\documentclass[prl,aps,twocolumn,showpacs]{revtex4-1}

\usepackage{graphicx} 
\usepackage{dcolumn}
\usepackage{bm}
\usepackage{amsmath}
\usepackage{amssymb}
\usepackage{color}
\usepackage{bbold}
\usepackage{bm}

\renewcommand{\vec}[1]{{\boldsymbol #1}}

\usepackage{natbib}

\newcommand{\llangle}[1][]{\savebox{\@brx}{\(\m@th{#1\langle}\)}%
  \mathopen{\copy\@brx\kern-0.5\wd\@brx\usebox{\@brx}}}
\newcommand{\rrangle}[1][]{\savebox{\@brx}{\(\m@th{#1\rangle}\)}%
  \mathclose{\copy\@brx\kern-0.5\wd\@brx\usebox{\@brx}}}

\begin{document}

\title{Bulk thermal transport coefficients in a quantum Hall system and the fundamental difference between thermal and charge response}
\smallskip

\author{Yuval Vinkler-Aviv}
\affiliation{University of Cologne, Institute for Theoretical Physics, 50937 Cologne, Germany}

\begin{abstract}
We derive and calculate thermal transport coefficient for a quantum Hall system in the linear response regime, and show that they are exponentially small in the bulk, in contrast to the quantized value of the charge Hall coefficient, thus violating Wiedemann-Franz law. This corroborates earlier reports about the essential difference between the charge and thermal quantum Hall effect, that originates from the different behavior of the corresponding $U(1)$ and gravitational anomalies. We explicitly calculate the bulk currents when a temperature profile is applied within the bulk, and show that they are proportional to the second derivative of the respective gravitational potential (tidal force), and nonuniversal, in contrast to the charge current which is proportional to the first derivative of the electrochemical potential.
\end{abstract}

\maketitle

{\em Introduction} -- The quantum Hall effect (QHE) is one of the most celebrated phenomenon in modern condensed matter physics, attracting theoretical and experimental attention since its discovery four decades ago~\cite{Klitzing_1980}. More recently, the thermal version of the QHE attracted interest, as it can supply additional unique information on the topological nature of the system under investigation, and sometimes it is the only signature of the QHE physics (in case that the excitations are neutral)~\cite{Kane_1997,Read_2000,Ryu_2012,Wang_2011,Stone_2012,Bradlyn_2015,Golan_2018}. A couple of recent experiments reported the measurement of a fractional quantum thermal Hall conductance~\cite{Banerjee_2017b,Banerjee_2018b,Kasahara_2018}, and spurred further theoretical investigations into the nature of this phenomenon~\cite{Ye_2018,Vinkler-Aviv_2018,Aharon-Steinberg_2019,Simon_2018,Feldman_2018,Ma_2019}.

Similar to the charge QHE, where a gradient in voltage leads to a perpendicular quantized electric current, in the quantized thermal QHE a temperature gradient leads to a universally valued heat current flowing in the perpendicular direction. The gapless chiral edges of the Hall bar carry quasiparticles and energy, and suggest a natural way to analyze and calculate both effects from the point of view of the edge. One of the hallmarks of the study of topological phases of matter is the bulk-edge correspondence (the holographic principle). In essence, bulk properties can be read-off of the edge properties, and vice-versa. In the charge QHE, the bulk-edge correspondence was beautifully explained by Laughlin's argument~\cite{Laughlin_1981}, which describes how a net charge flows between the edges through the gapped bulk, in response to a force induced by a change in the magnetic field. The argument relates the magnetic flux to the net charge transported, and derives the quantized Hall conductance from this relation.

The relation between the edge and the bulk can also be understood through the concept of quantum anomalies~\cite{Witten_1984,Cappelli_2002,Fujikawa2004path}. The theory that describes the edge of the QHE is an anomalous one, which reflects the fact that it is an effective theory that lies on the boundary of a higher dimensional one. The anomalies are manifested in the nonconservation of currents along the edge. The theoretical tool of anomalies was used to generalize and classify topological theories and their possible transport signatures~\cite{Schnyder_2008,Ryu_2010,Wang_2011,Ryu_2012}.

In contrast to charge transport, where one can couple the particle density to an electrochemical potential and derive the transport coefficients from the response, there is no apparent way to add to the Hamiltonian a term that describes the temperature. A trick developed by Luttinger~\cite{Luttinger_1964} allowed the calculation of thermal transport coefficient by coupling the system to gravity. As both elements similarly affect the energy density, the response of the system to a weak gravitational potential is identical to its response to small temperature gradient. Consequently, the anomaly that is used in order to describe the thermal QHE is a gravitational anomaly, and through it one can classify the topological classes to which different systems belong.

This straight-forward analogy between charge and thermal QHE was qualified in a work by Stone~\cite{Stone_2012}, who outlined a central distinction between the two effects, based on the study of the nature of the respective anomalies. Thermal currents in the linear regime, according to Stone, can only flow along the gapless edges, and no transport of heat is possible through the gapped bulk. A bulk response to temperature requires higher order gradients of the gravitational potential, that create tidal forces. Several works in recent years have approached the issue from different angles~\cite{Bradlyn_2015,Nakai_2016,Nakai_2017,Golan_2018}, considering different forms of perturbation and anomalies. The ongoing theoretical debate over the nature of the thermal QHE is ever more relevant in light of the recent experiments, that raised questions regarding the relationship between the edge and the gapped bulk. Specifically, it has been suggested that gapless bulk phonons, which allow thermal conductance down to the lowest temperatures, play a central role in enabling the measurement of the thermal QHE and transport heat between the gapless edges through the bulk~\cite{Vinkler-Aviv_2018,Ye_2018}.

In this letter we present a direct calculation of the bulk thermal transport coefficients in a quantum Hall system, based upon a model for a topological insulator. We show that while the model present a quantized $\sigma_{xy}$ its thermal Hall conductivity $\kappa_{xy}$ is exponentially small in $M/T$ where $M$ is the bulk energy gap. This breaks Wiedemann-Franz law in the topological state, in contrast to the topologically trivial regime in which the law is maintained. We further calculate explicitly the bulk thermal Hall response to a space dependent perturbation, and show that it is higher order in the spatial derivative than the charge response, corresponding to tidal forces.

{\em Model and linear response calculations} -- We consider a standard continuum model describing a topological insulator~\cite{Hasan_2010}, given by the Hamiltonian $\mathcal{H} = \int\! d^2k \Psi^{\dagger}_{\vec{k}} h(\vec{k})\Psi_{\vec{k}}$ with $\Psi_{\vec{k}}$ a 2-component spinor and
\begin{equation}
	h(\vec{k}) = (M+\lambda^2 \vec{k}^2)\sigma_z +v(k_x \sigma_x + k_y \sigma_y)
\end{equation}
The gap at $\vec{k}=0$ is $M$, and its sign defines whether the system is topologically trivial ($M>0$) or not ($M<0$). In the non-trivial phase the model is characterized by a Chern number of $-1$. This model can be derived by expanding about a Dirac point on the surface of a topological insulator or a superconductor, and we later consider a concrete lattice model. Our goal here is to calculate the thermal transport coefficients for this model, and specifically derive the Hall thermal conductivity $\kappa_{xy}$.

An appealing route to derive bulk thermal transport coefficient is by employing Wiedemann-Franz law~\cite{Smrcka_1977,Nasu_2015,Nasu_2017}, based, for example, on the Boltzmann scattering picture. Assuming that heat is carried in a similar manner to charge, by scattering of quasiparticles, one can relate the thermal and particle (charge) Hall conductivity by
\begin{equation}
	\kappa_{xy} = -\frac{1}{e^2 T}\int\! d\epsilon (\epsilon-\mu)^2 \sigma_{xy}
				(\epsilon)f'(\epsilon).
	\label{eq:kxy_from_WF}
\end{equation}
Here, $\mu$ is the chemical potential relative to which the energy of excitations is measured, $e$ is their charge, which we shall set to $1$ from henceforward, $f(\epsilon)$ the Fermi-Dirac distribution and $\sigma_{xy}(\epsilon)$ is the charge Hall conductivity at zero temperature and at chemical potential $\epsilon$, that can be calculated from the standard Kubo formula. However, this straight-forward application of the Wiedemann-Franz law might fail if the transport mechanisms for charge and for heat are essentially different. This indeed is the case in the topological regime, as we shall see here.

In order to directly calculate the thermal transport coefficients for this model, we first need to derive an expression for the local thermal current $\hat{\vec{j}}^Q(\vec{r})$. This can be done by coupling the Hamiltonian to a fictitious gravitational potential $\psi(\vec{r})$ (not to be confused with the fermionic fields $\Psi$), such that it is given by $\mathcal{H}_{\psi} = \int\!d^2r [1+\psi(\vec{r})]\hat{h}(\vec{r})$. Now, we require both that the current will satisfy the continuity equation $\vec{\nabla}\hat{\vec{j}}^Q_\psi(\vec{r}) = i[1+\psi(\vec{r}][\hat{h}(\vec{r}),\mathcal{H}_\psi]$ and that it will scale locally with the potential $\hat{\vec{j}}^Q_\psi(\vec{r}) = [1+\psi(\vec{r})]^2\hat{\vec{j}}^Q_{\psi=0}(\vec{r})$~\cite{Cooper_1997,Qin_2011,Bradlyn_2015}. After the derivation we can set $\psi=0$ for the heat current in a flat background. Carrying out this calculation~\cite{sup_mat} we derive the following expression for the thermal current operator in momentum basis $\hat{\vec{j}}^Q_{\vec{q}}=\int\!d^2k \Psi^{\dagger}_{\vec{k}}\vec{j}^Q_{\vec{k},\vec{k}-\vec{q}}\Psi_{\vec{k}-\vec{q}}$, with
\begin{widetext}
\begin{eqnarray}
	\vec{j}^Q_{\vec{k},\vec{k}'} &=& (v^2+2M\lambda^2 +
	2\lambda^4 \vec{k}\cdot\vec{k}')
	\frac{\vec{k}+\vec{k}'}{2}
	+iv\lambda^2[(\vec{k}+\vec{k}')\times\vec{\sigma}]\cdot\hat{z}
	\frac{\vec{k}-\vec{k}'}{2} 
	\nonumber \\ &&
	-i[v^2\sigma_z-\lambda^2
	(2i\lambda^2(\vec{k}\times\vec{k}')_z+v(\vec{k}+\vec{k}')\cdot
	\vec{\sigma})]\hat{z}\times\frac{\vec{k}-\vec{k}'}{4}.
	\label{eq:j_kkp}
\end{eqnarray}
\end{widetext}
The charge current is given similarly by $\hat{\vec{j}}^N_{\vec{q}}=\int\!d^2k \Psi^{\dagger}_{\vec{k}}\vec{j}^N_{\vec{k},\vec{k}-\vec{q}}\Psi_{\vec{k}-\vec{q}}$ with
\begin{equation}
	\hat{\vec{j}}^N_{\vec{k},\vec{k}'} = v\sigma_x \hat{x} + v\sigma_y \hat{y}
					+\lambda^2(\vec{k}+\vec{k}')\sigma_z.
\end{equation}

As time reversal symmetry is broken, there are magnetization currents in the sample even at thermal equilibrium. These currents are included in the expression for $\hat{\vec{j}}^Q(\vec{r})$ but do not contribute to the net transport of energy that is measured by $\kappa_{xy}$, as was discussed in detail in Refs.~\cite{Cooper_1997,Qin_2011}. Therefore, the correct calculation of $\kappa_{xy}$ includes two contributions $\kappa_{xy} = \kappa_{xy}^{\rm Kubo} + \kappa_{xy}^{M}$ where
\begin{eqnarray}
	\kappa_{xy}^{\rm Kubo} &=& \frac{L^{22}_{xy}}{T^2} = 
	\frac{1}{T}\int\! dt e^{-\eta t}\langle \hat{J}^Q_x(0);\hat{J}^Q_y(t)\rangle,
	\nonumber \\
	\kappa_{xy}^{M} &=& \frac{2M^Q_z}{T}.
\end{eqnarray}
Here $\langle \hat{A} ; \hat{B} \rangle = T\int_0^{\beta}\! d\lambda \langle \hat{A}(-i\lambda)\hat{B}\rangle$ is the Kubo correlation function, and the magnetization quantity $\vec{M}^Q$ is defined by the differential equation
\begin{equation}
	2\vec{M}^Q - T \frac{\partial \vec{M}^Q}{\partial T} = 
	-\frac{i}{2T}\vec{\nabla}_\vec{q} \times
	\langle \hat{h}_{-\vec{q}} ; \hat{\vec{j}}^Q_\vec{q} \rangle
	|_{\vec{q}\to 0}
\end{equation}
with the boundary condition that at zero temperature $2\vec{M}^Q$ coincides with the right hand side.

At temperatures well below the gap $T \ll |M|$, a correlation function between two operators will not be exponentially small in $|M|/T$ only if {\it both} operators include matrix elements between states in the conductance and valence bands, that is -- terms which are off-diagonal in the energy basis. As at the limit $\vec{q}\to 0$ both $\hat{\vec{j}}^Q_{\vec{q}}$ and $\hat{h}_{\vec{q}}$ do not contain any such off-diagonal elements, $\kappa_{xy}$ will be exponentially small in $|M|/T$ for $T \ll |M|$, regardless of the sign of $M$. This, in contrast with the expression for the charge current which has $\sigma_j$ components, and its correlations attain the quantized value of $\sigma_{xy}=-G_0$ for $M<0$ at low temperatures.

\begin{figure}[t]
\begin{tabular}{cc}
\includegraphics[width=0.23\textwidth]{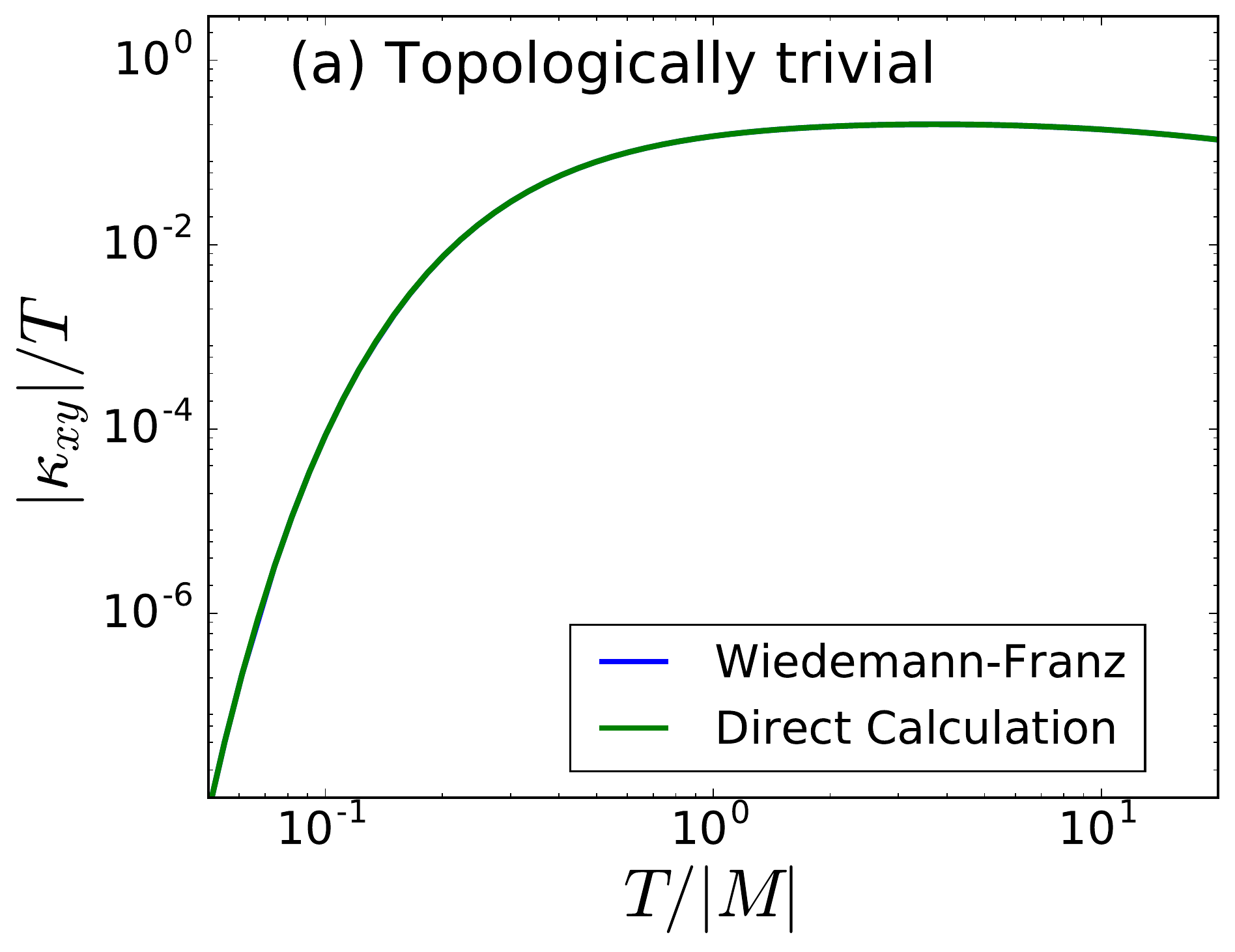}
&
\includegraphics[width=0.23\textwidth]{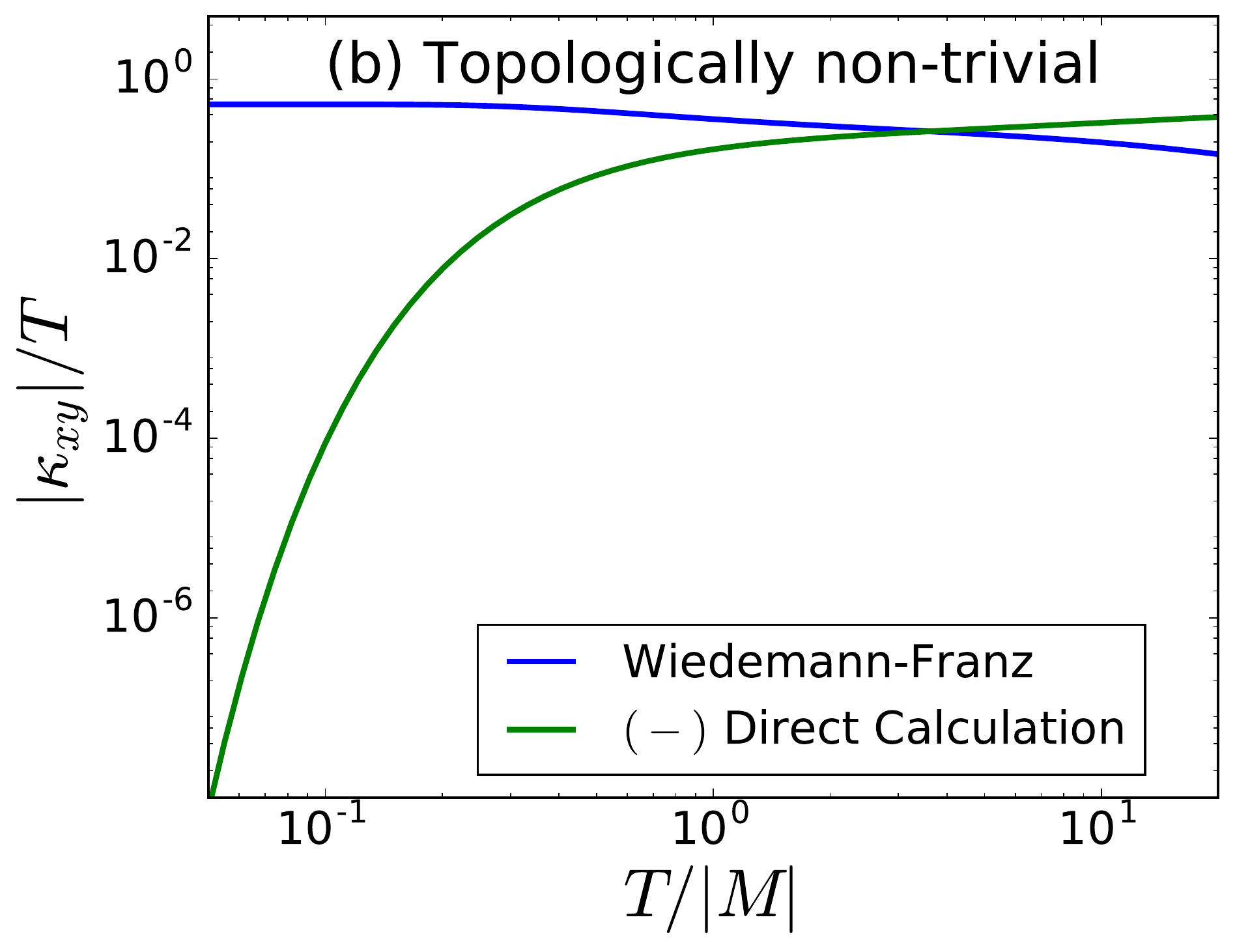}
\end{tabular}
\caption{(color online) Comparison of the calculation of $\kappa_{xy}$ as a function of temperature using Wiedemann-Franz law of Eq.~(\ref{eq:kxy_from_WF}), against its direct calculation from the transport coefficients. The two methods agree in the topological trivial case (a), but differ significantly for the non-trivial case with $M<0$ (b). Note that in the topological non-trivial case the results not only differ in magnitude but also in the sign of $\kappa_{xy}$. In all calculations here $\lambda^2=10^{-2}v^2/|M|$.}
\label{fig:kxy_WF}
\end{figure}

An explicit calculation of the different correlation functions~\cite{sup_mat} gives the following results
\begin{eqnarray}
	\kappa_{xy}^{\rm Kubo} &=& 0, \nonumber \\
	\kappa_{xy}^{M} &=& -
	\int_0^{\infty}\! \frac{dk}{2\pi} \frac{\lambda^2 k^2-M}{E_{k}^3}v^2 k
	\bigg\{T{\rm Li}_2\left(\!-e^{-\tfrac{E_k}{T}}\right)
	\nonumber \\ && +
	E_k\ln\left[f(-E_k)\right]-\frac{E_k^2}{2T}f(E_k)
	\bigg\},
	\nonumber \\
	\sigma_{xy} &=& \int_0^{\infty}\! \frac{dk}{2\pi}
	\frac{\lambda^2 k^2-M}{2 E_{k}^3}v^2 k
	\left[f(E_{k})-f(-E_{k})\right],
\end{eqnarray}
where $E_k = \sqrt{(\lambda^2k^2+M)^2+v^2k^2}$ and ${\rm Li}_2(x)$ is the second-order polylogarithm function. In Fig.~(\ref{fig:kxy_WF}) we compared the direct calculation of $\kappa_{xy}$ with the one resulting by relying on Wiedemann-Franz law. In the topological trivial phase $(M>0)$, where no QHE exist and Laughlin's argument does not apply, the two calculations agree. However, in the QHE phase ($M<0$) Wiedemann-Franz law breaks, as the charge transport mechanism is different than the thermal one, and it gives an incorrect non-vanishing, universal value of $\kappa_{xy}/T=\pi k_B^2/6\hbar$ at $T\ll -M$. It is important to point out that the inclusion of gapless chiral edge states that connect both edges, will allow thermal conductance at low temperatures, with the correct universal value, and restore Wiedemann-Franz law~\cite{Bradlyn_2015}. The universal value might also be restored when other gapless modes of transferring heat between the edges exist, such as bulk phonons that couple to the edge modes, as was pointed in two recent studies~\cite{Vinkler-Aviv_2018,Ye_2018}.

{\it Tidal forces} -- While the bulk linear response to small temperature gradients is exponentially suppressed, the anomaly is activated when tidal gravitational forces, originating from spatial derivatives of the gravitational potential higher than the linear order, are present~\cite{Stone_2012,Golan_2018}. This can be seen by calculating the bulk heat Hall currents under a space-dependent gravitational potential. The equivalent charge setup will describe bulk charge Hall currents under a varying electrochemical potential. This is a known phenomena in Hall systems, where local gates in the bulk create perpendicular dissipationless currents around them.

To calculate such responses, we place our model on a finite lattice of $N_x \times N_y$ sites. Now we can break translation invariance in the $x$-direction by considering a space dependent potential, which is zero near the boundaries and finite deep inside the bulk, and explicitly calculate the perpendicular induced current, taking advantage of the fact that the model is quadratic and can diagonalized exactly. To avoid edge effects, we placed the system on a torus by enforcing periodic boundary conditions.

The lattice model Hamiltonian is given by $\mathcal{H}=\sum_{\vec{k}}\Psi^{\dagger}_{\vec{k}}h_\vec{k}\Psi_{\vec{k}}$ with
\begin{equation}
	h_{\vec{k}}=\frac{v}{a}\sum_{j=x,y}\sin(k_j a)\sigma_j+
	\left[M-2\frac{\lambda^2}{a^2}\sum_{j=x,y}\cos(k_j a)
	\right]\sigma_z,
\end{equation}
where $a$ is the lattice spacing. This lattice Hamiltonian is topologically nontrivial for $|Ma^2/\lambda^2| < 4$ with a Chern number of ${\rm sgn}\{M\}$ in these regimes. The gap closures at $M=0, M=\pm 4\lambda^2/a^2$ denote topological phase transitions~\cite{Qi_2006}. We choose the local Hamiltonian density to be written as
\begin{equation}
	\hat{h}_{\rm lat}(\vec{r}) = \frac{1}{N_x N_y}\sum_{\vec{k},\vec{k}'}
	e^{-i(\vec{k}-\vec{k}')\vec{r}}
	\Psi^{\dagger}_{\vec{k}}\vec{d}_{\rm lat}(\vec{k},\vec{k}')\cdot\vec{\sigma}
	\Psi_{\vec{k}'},
\end{equation}
with
\begin{eqnarray}
	\vec{d}_{\rm lat}&=&
	\sum_{j=x,y}
	\frac{v}{a}e^{i(k_j-k'_j)\tfrac{a}{2}}
	\sin\left(\frac{k_j+k'_j}{2}a\right)\hat{r}_j+
	\nonumber \\ &&
	\left[M-
	2\frac{\lambda^2}{a^2}
	\sum_{j=x,y}
	e^{i(k_j-k'_j)\tfrac{a}{2}}
	\cos\left(\frac{k_j+k'_j}{2}a\right)
	\right]\hat{z}.
	\nonumber \\ &&
\end{eqnarray}

We derive the heat current in the $y$-direction at on the lattice at each point $\hat{j}^Q_y(x,y)$~\cite{sup_mat}, and then calculate
\begin{eqnarray}
	j^Q_y(\vec{r}) &=& \langle \hat{j}^Q_y(\vec{r})\rangle_{\psi}-
	\langle \hat{j}^Q_y(\vec{r})
	\rangle_{\psi=0} 
	\nonumber \\ &=& {\rm Tr}\left[\left(\hat{\rho}_{\psi}-
	\hat{\rho}_{\psi=0}\right)
	\hat{j}^Q_y(\vec{r})\right]
\end{eqnarray}
with $\hat{\rho}_{\psi}$ the density matrix described by a space-dependent Hamiltonian of the form $\mathcal{H}_{\psi,{\rm lat}} = \sum_{\vec{r}}[1+\psi(x)]\hat{h}_{\rm lat}(\vec{r})$, and we deduct the equilibrium expectation value of the energy currents, that exist even in the absence of temperature gradients.
	
One should distinguish between thermal currents in response to temperature gradients and energy currents in response to a true gravitational (or geometrical) field that changes the Hamiltonian, such as calculated for example in Ref.~\cite{Golan_2018}. Luttinger's argument pertained to the {\it state} of the system, and his observation was that the density matrix in response to a small fictitious gravitational field is identical to the one when a small temperature gradient is applied~\cite{Luttinger_1964,Stone_2012}. However, when true gravitation is applied the {\it definition} of energy is also changed, and the operator that describe the energy current will scale as $\hat{\vec{j}}^Q_{\psi}(\vec{r}) = [1+\psi(\vec{r})]^2\hat{\vec{j}}^Q_{\psi=0}(\vec{r})$. When temperature gradients are considered, the definition of the energy is not affected, and therefore the correct operator describing the heat current is $\hat{\vec{j}}^Q_{\psi=0}$, and its expectation value should be taken with respect to $\mathcal{H}_{\psi}$. The difference between the two approaches is already in the linear order in the potential, as can be seen by considering a gravitational potential of the form $1+\epsilon \psi(\vec{r})$ and expanding to leading order in $\epsilon$. The density matrix is then $\hat{\rho}_\psi \simeq \hat{\rho}_{0}+\epsilon \delta\hat{\rho}$ and the different responses are
\begin{eqnarray}
	\vec{j}_{T}^Q(\vec{r}) &\simeq & 
	\epsilon{\rm Tr}\left[\delta\hat{\rho} \hat{\vec{j}}^Q_0\right],
	\nonumber \\ 
	\vec{j}_{\rm g}^Q(\vec{r}) &\simeq & 
	2\epsilon \psi(\vec{r})\vec{j}^Q_0 +
	\epsilon{\rm Tr}\left[\delta\hat{\rho} \hat{\vec{j}}^Q_0
	\right],
\end{eqnarray} 
where $\vec{j}^Q_{\rm g}$ is the energy current in response to true gravitational potential and $\vec{j}^Q_{T}$ in response to temperature gradients. If time-reversal symmetry is not broken and there are no energy currents in equilibrium $\vec{j}^Q_0=0$, then the two calculations agree to leading order.

\begin{figure}[t]
\begin{tabular}{cc}
\includegraphics[width=0.23\textwidth]{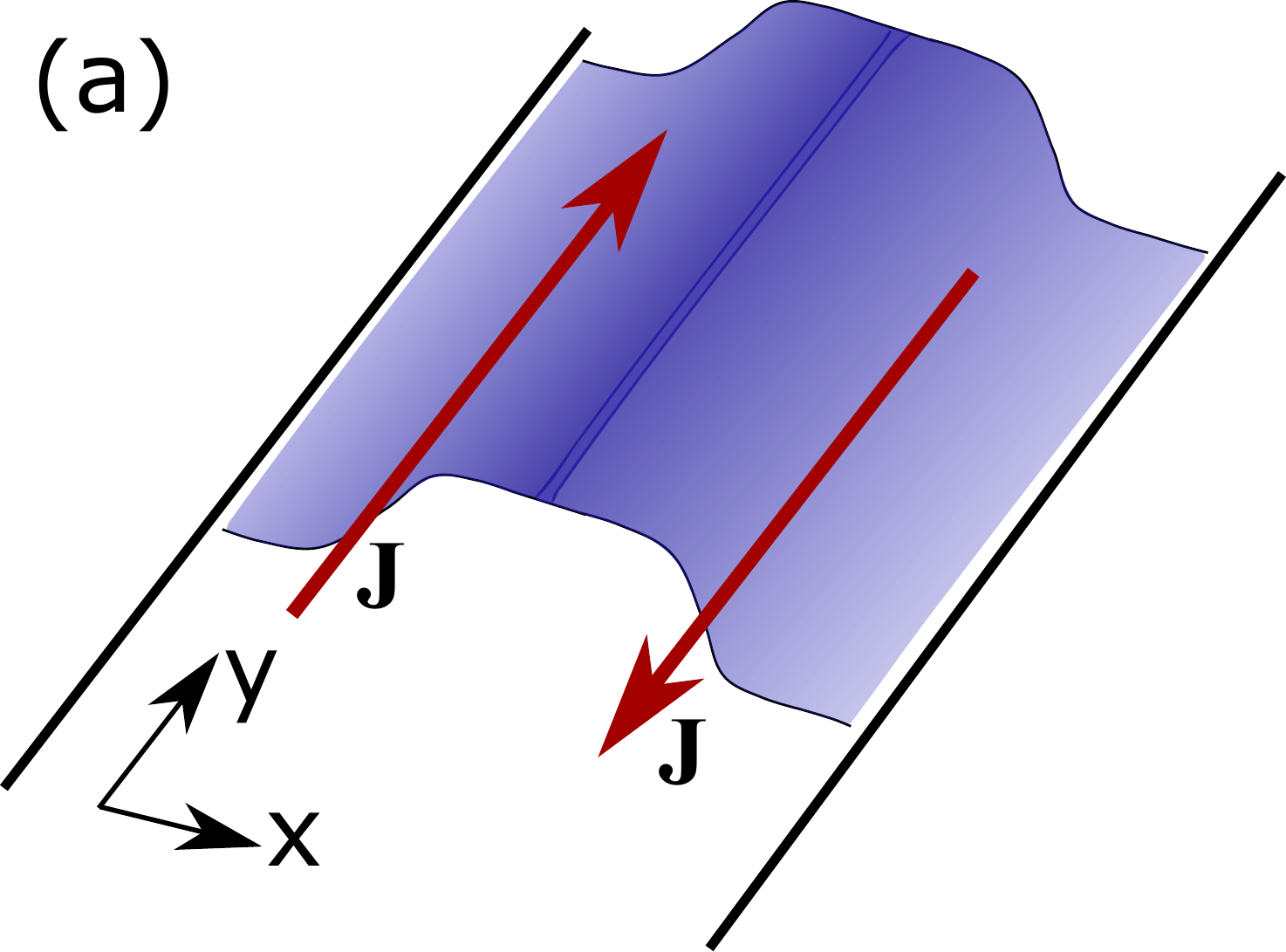}
&
\includegraphics[width=0.23\textwidth]{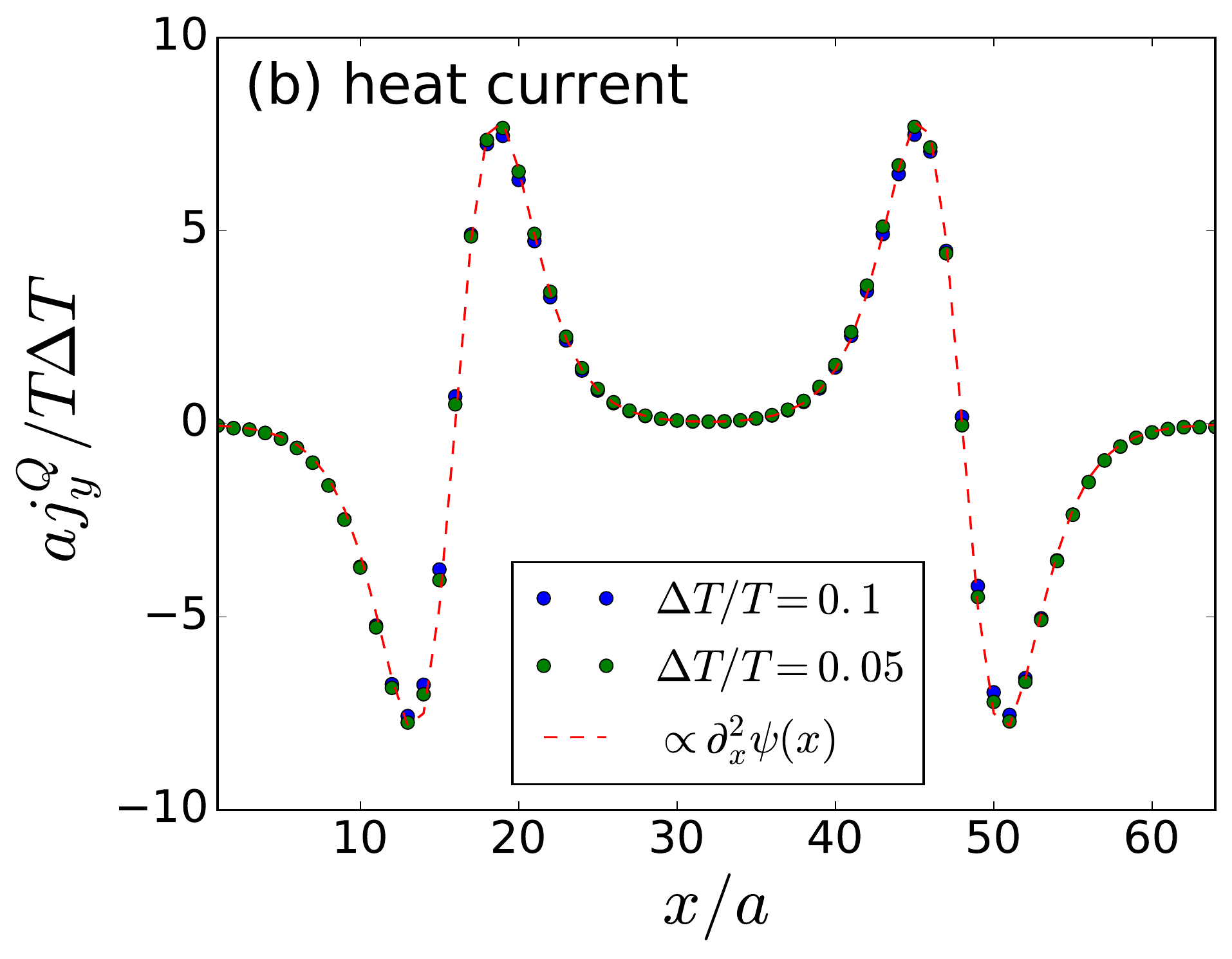}
\\
\includegraphics[width=0.23\textwidth]{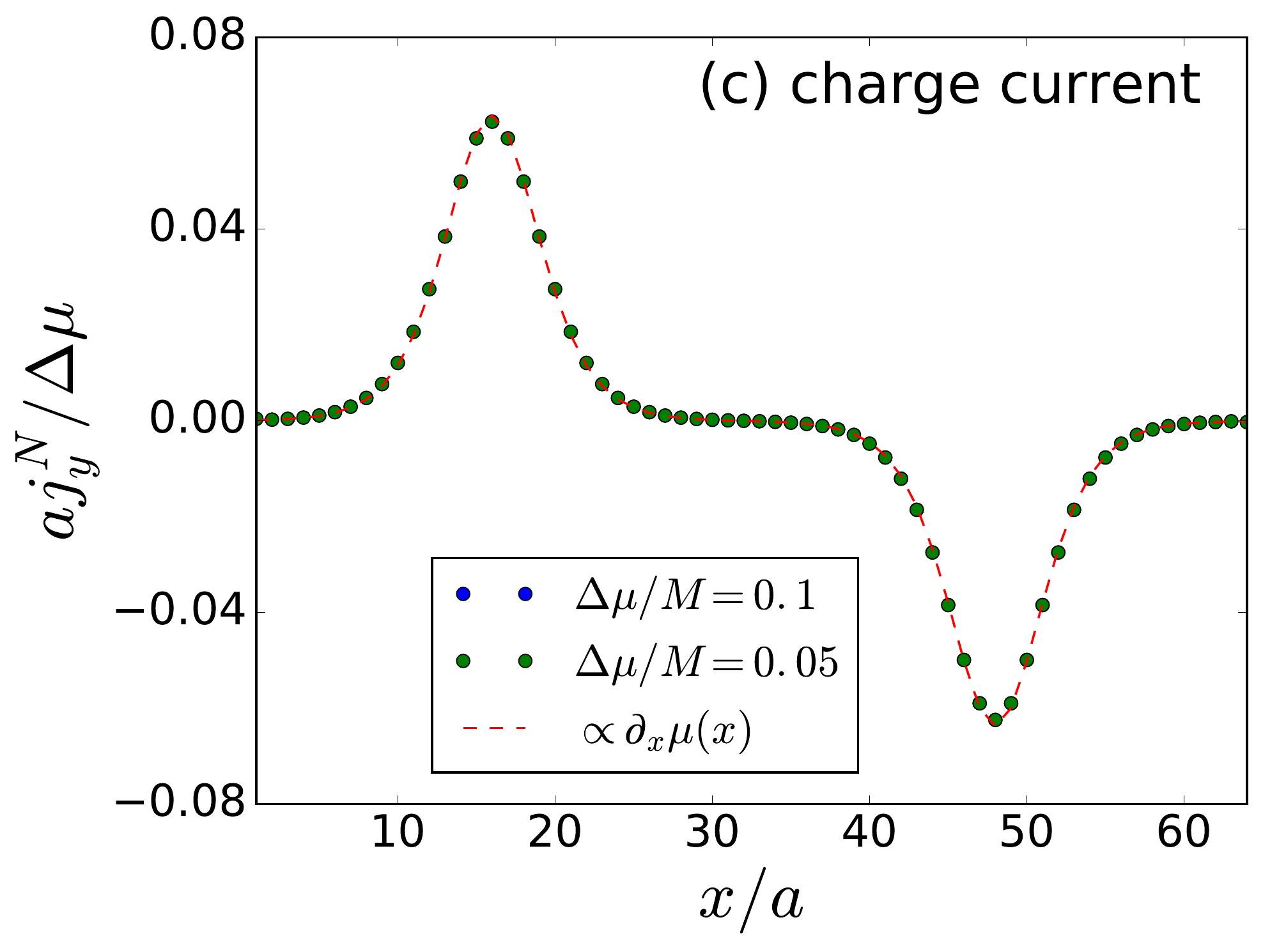}
&
\includegraphics[width=0.23\textwidth]{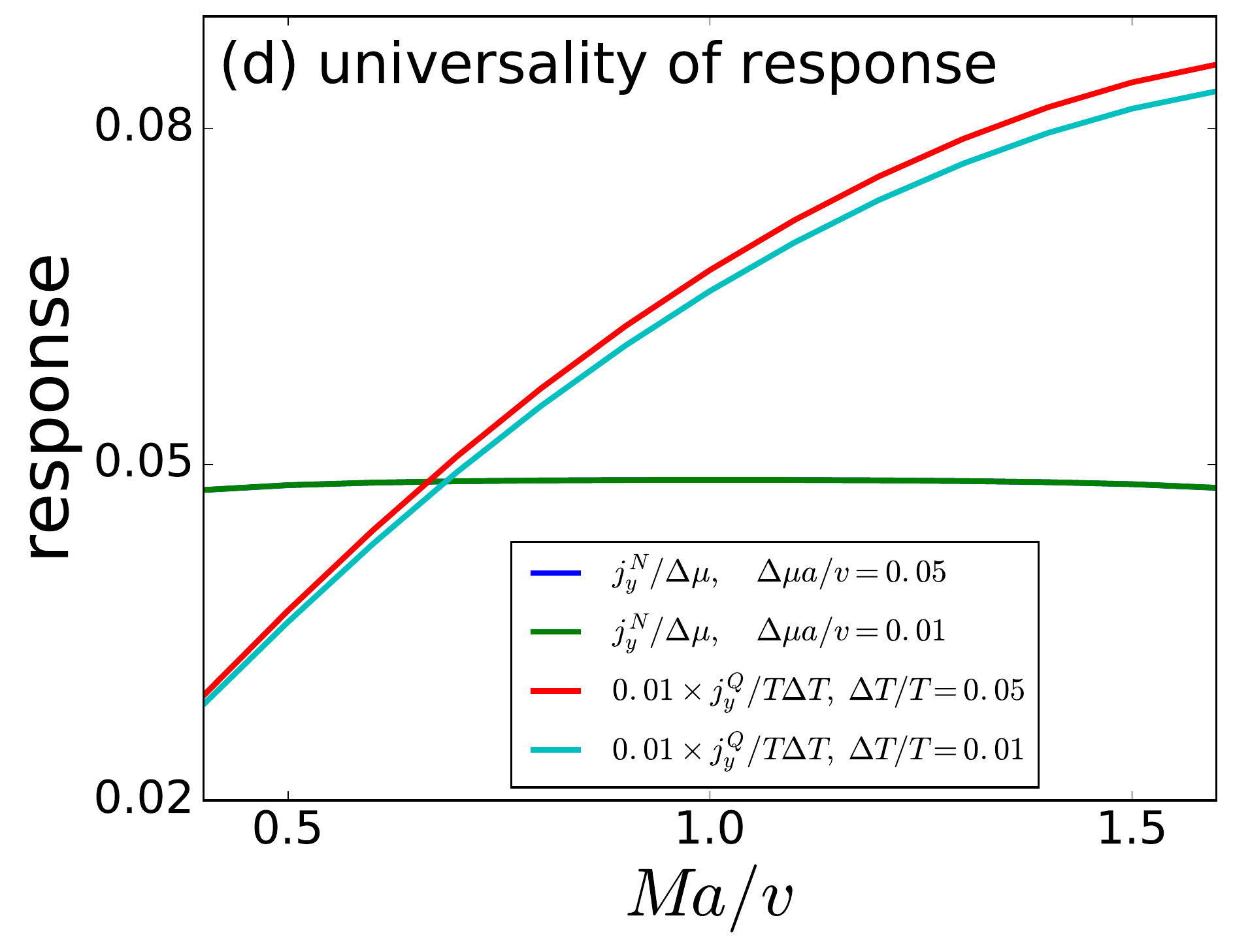}
\end{tabular}
\caption{(color online) Hall currents in the bulk, for the setup depicted in (a), where the system lies on a finite lattice with translation invariance maintained in the $y$ direction and space dependent potential in the $x$-direction. In (b) we plotted the thermal Hall currents in the $y$ direction at different points, in response to a gravitational potential $\psi(x) = (\Delta T/T)f(x)$, and in (c), for comparison, the charge response to to an electrochemical potential $\mu(x)=\Delta\mu f(x)$, where $f(x)$ is as given in Eq.~(\ref{eq:smooth_potential}). We superimposed the thermal (charge) currents with the second (first) spatial derivative of the potential, in dashed red lines, scaled to match. In all calculations the lattice had $N_x=64, N_y=128$ sites, and the parameters used were $x_L=16a$, $x_R=48a$, $\xi=4a$, $M=v/a$, $2(\lambda/a)^2=M$, and $T=M/20$. In (d) we plot the dependence of the thermal (charge) current at a fixed point $x=20a$ ($x=18a$) as a function of $M$, for $2(\lambda/a)^2=v/a$, $T=v/20a$ and $f(x)$ identical to the one used in $(b),(c)$.}
\label{fig:lat_cur}
\end{figure}

We choose $\psi(x) = (\Delta T/T)f(x)$ with
\begin{equation}
	f(x) = 	\frac{1}{2}\left[ \tanh\left(\frac{x-x_L}{\xi}\right)
	-\tanh\left(\frac{x-x_R}{\xi}\right)\right],
	\label{eq:smooth_potential}
\end{equation}
such that it changes smoothly from zero near the edges to a uniform value in the bulk, see Fig.~(\ref{fig:lat_cur}a). For comparison, we also calculated the bulk charge current $\langle \hat{j}^N_y(\vec{r})\rangle_{\mu}$ in response to a space-dependent chemical potential $\mu(\vec{r})$ that we add to the Hamiltonian $\mathcal{H}_{\mu,\rm lat} = \sum_{\vec{r}}[\hat{h}_{\rm lat}(\vec{r})+\mu(x)\hat{n}(\vec{r})]$, where $\hat{n}(\vec{r})$ the local  particle density.

The results of the calculations are given in Fig.~(\ref{fig:lat_cur}b-c). The charge Hall current at low temperatures $T\ll|M|$ is proportional to the first derivative of the potential $j^N_y(x) \propto \partial_x \mu(x)$, as expected. However, the thermal Hall current requires tidal forces, and is proportional to the \emph{second} derivative of the gravitational potential $j_y^Q(x,0)\propto \partial^2_x \psi(x)$. Furthermore, while the coefficient of the charge current is the universal quantized value, the heat current response is not~\cite{Golan_2018}, as can be seen in Fig.~(\ref{fig:lat_cur}d), where we plotted the Hall current at a fixed point when changing $M$.  This corroborate the exponentially vanishing linear response to temperature gradients, and further clarifies the different nature of the thermal and charge responses.

{\em Thermopower and Onsager relations} -- Finally, we consider the bulk thermopower properties. That is the charge currents in response to temperature gradients and thermal currents in response to electrochemical potential gradients. This is of interest for two reasons. The first being that measurement of charge currents is much more accessible experimentally than measurement of heat currents. The other is due to Onsager relations that we expect to be maintained, where these two responses are similar.

\begin{figure}[t]
\includegraphics[width=0.43\textwidth]{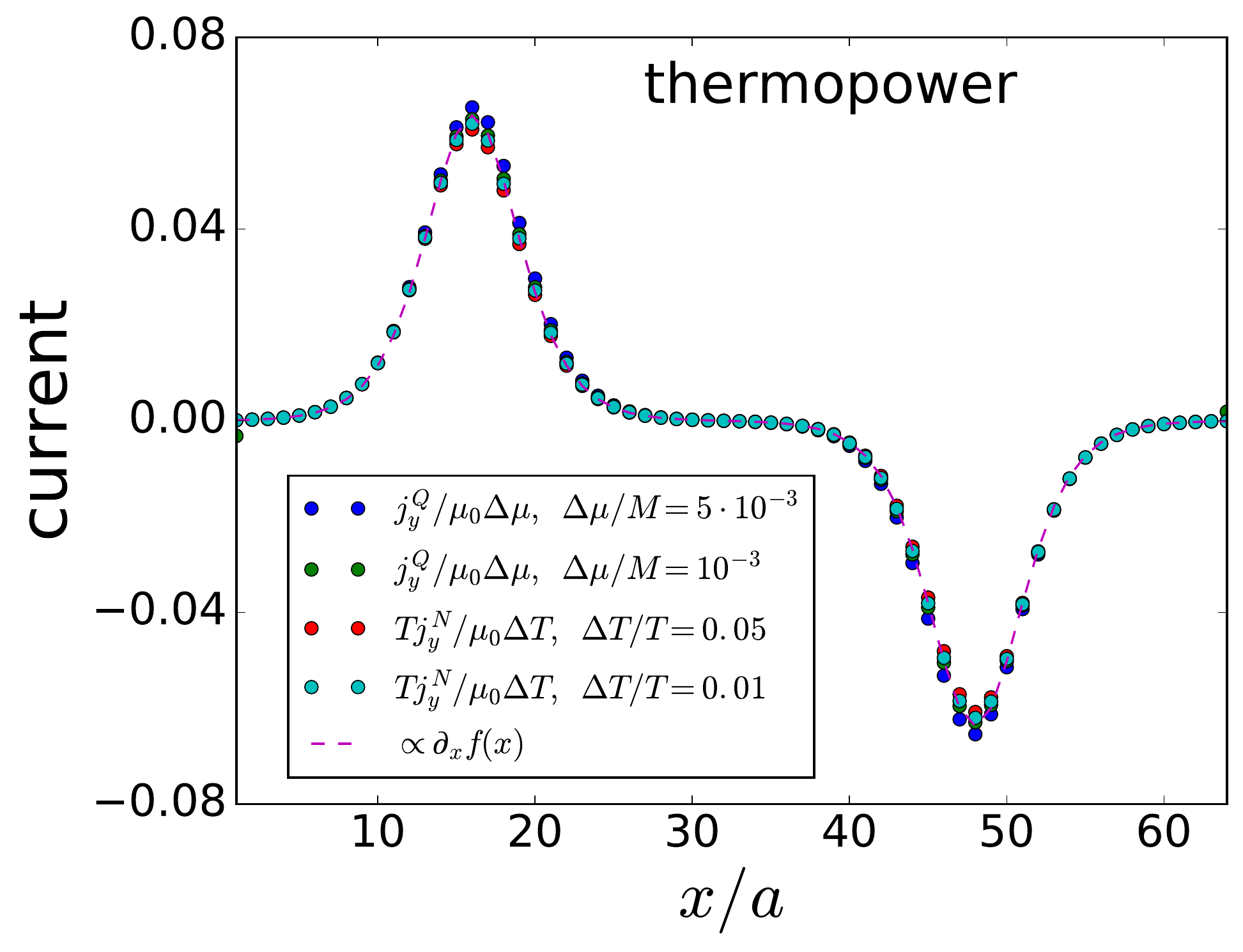}
\caption{(color online) Thermopower currents in the bulk, including both heat current in response to space-dependent electrochemical potential and charge current in response to space-dependent temperature. All scale with $\mu_0$ that breaks particle-hole symmetry, and will vanish in its absence. The parameters used here are identical to the ones used in Fig.~(\ref{fig:lat_cur})(b-c) with the addition of $\mu_0/M = 0.05$}
\label{fig:thermopower}
\end{figure}

In order to get finite charge response to temperature gradients one has to break the model's particle-hole symmetry, for example by adding a constant chemical potential $\mathcal{H}_{\psi,{\rm lat}} \to \mathcal{H}_{\psi,{\rm lat}}+\mu_0 \sum_{\vec{r}}\hat{n}(\vec{r})$, which by itself does not induce any type of current. However, one needs to adjust the expression for the heat current accordingly, as now it is given by $\hat{\vec{j}}^Q_{\mu} = \hat{\vec{j}}^Q_{\mu=0}+\mu \hat{\vec{j}}^N$. Examining the induced thermal (charge) currents when chemical potential (temperature) is varied spatially [see Fig.~(\ref{fig:thermopower})], they indeed follow similar behavior, and both are linear in the spatial derivative of the perturbation.

{\it Discussion} -- Analyzing and calculating bulk thermal currents in the QHE, we explicitly showed that the linear response thermal transport coefficients in the {\it bulk} are exponentially small $|M|/T$, leading to a violation of Wiedemann-Franz law. This demonstrates that the bulk Hall conductance for charge and for energy is fundamentally different. To further corroborate this, we explicitly calculated how do bulk thermal Hall currents behave when small space-dependent perturbation is applied in the bulk. While charge currents are proportional to the first space derivative of the perturbation, with a universal coefficient, the thermal currents are proportional to the its second derivative, with nonuniversal coefficient. This means that tidal forces are necessary in order to induce bulk thermal Hall currents, and further stresses the difference between the two phenomena, that as Stone showed can be traced back to the different type of anomalies causing them. However, the thermopower relations are universal, and should be measurable in an experiment where temperature gradient causes quantized electrical Hall currents to flow, which can be detected.

{\it Acknowledgments} -- The authors would like to thank A.\ Altland, A.\ Stern, O.\ Golan, V.\ Dwivedi, C.\ Hickey and especially A.\ Rosch for useful discussions and comments. This work was supported by the DFG (project C02 of CRC1283 and project A01 of CRC/TR183).

\bibliography{kxy_bib}

\begin{widetext}

\section{Supplementary Material}
\nonumber 

\subsection{Derivation of the heat current}
In order to derive the expression for the heat current, we start by writing down the continuity equation in presence of the gravitational potential $\psi(x)$
\begin{eqnarray}
	\vec{\nabla}\vec{j}_\psi^Q(\vec{r}) &=& i\left[1+\psi(\vec{r}\right]
	\left[ \hat{h}(\vec{r}),\mathcal{H}_{\psi}\right]=
	i[1+\psi(\vec{r})]
	\int\! \frac{d^2k d^2k' d^2q}{(2\pi)^4}d^2r'[1+\psi(\vec{r}')] \times
	\nonumber \\ &&
		\Psi^{\dagger}_{\vec{k}}
	\left[\vec{d}(\vec{k},\vec{q})\cdot\vec{d}(\vec{q},\vec{k}')+
	i\vec{d}(\vec{k},\vec{q})\times\vec{d}(\vec{q},\vec{k}')\cdot\vec{\sigma}
	\right]
	\left[e^{-i(\vec{k}-\vec{q})\vec{r}-i(\vec{q}-\vec{k}')\vec{r}'}-
	e^{-i(\vec{q}-\vec{k}')\vec{r}-i(\vec{k}-\vec{q})\vec{r}'}\right]
	\Psi_{\vec{k}'},
\end{eqnarray}
where
\begin{equation}
	\vec{d}(\vec{k},\vec{k}') = v\frac{\vec{k}+\vec{k}'}{2}+
	\left[M+\lambda^2 \vec{k}\cdot\vec{k}'\right]\hat{z},
\end{equation}
is a $3$-dimensional vector in band (spin) space, and $\vec{k},\vec{k}', \vec{q}$ are $2$-dimensional vectors in real space. In order to make the expression local, we (i) replace $\vec{q}$ inside the different $\vec{d}$ with derivatives with respect to $\vec{r}'$
\begin{eqnarray}
	\vec{q} e^{-i(\vec{k}-\vec{q})\vec{r}-i(\vec{q}-\vec{k}')\vec{r}'} &=& 
	\left(i\vec{\nabla}_{\vec{r}'}+\vec{k}'\right)
	e^{-i(\vec{k}-\vec{q})\vec{r}-i(\vec{q}-\vec{k}')\vec{r}'},
	\nonumber \\ 
	\vec{q} e^{-i(\vec{k}-\vec{q})\vec{r}'-i(\vec{q}-\vec{k}')\vec{r}} &=& 
	\left(-i\vec{\nabla}_{\vec{r}'}+\vec{k}\right)
	e^{-i(\vec{k}-\vec{q})\vec{r}'-i(\vec{q}-\vec{k}')\vec{r}},
\end{eqnarray}
and then (ii) use integration by parts in order to move the derivative to act on $\left[1+\psi(\vec{r}')\right]$. Now we can (iii) integrate over $\vec{q}$, getting $\delta(\vec{r}-\vec{r}')$ and making the expression local, with derivatives of the gravitational potential
\begin{eqnarray}
	\vec{\nabla}\vec{j}_\psi^Q(\vec{r}) &=&
	i[1+\psi(\vec{r})]
	\int\! \frac{d^2k d^2k'}{(2\pi)^2}e^{-i(\vec{k}-\vec{k}')\vec{r}}
	\Psi^{\dagger}_{\vec{k}}
	\bigg[\vec{d}(\vec{k},-i\vec{\nabla}_{\vec{r}}+\vec{k}')
	\cdot\vec{d}(-i\vec{\nabla}_{\vec{r}}+\vec{k}',\vec{k}')
	\nonumber \\ && +
	i\vec{d}(\vec{k},-i\vec{\nabla}_{\vec{r}}+\vec{k}')\times
	\vec{d}(-i\vec{\nabla}_{\vec{r}}+\vec{k}',\vec{k}')\cdot\vec{\sigma}
	-\vec{d}(\vec{k},i\vec{\nabla}_{\vec{r}}+\vec{k})
	\cdot\vec{d}(i\vec{\nabla}_{\vec{r}}+\vec{k},\vec{k}')
	\nonumber \\ &&
	-i\vec{d}(\vec{k},i\vec{\nabla}_{\vec{r}}+\vec{k})\times
	\vec{d}(i\vec{\nabla}_{\vec{r}}+\vec{k},\vec{k}')\cdot\vec{\sigma}
	\bigg][1+\psi(\vec{r})]\Psi_{\vec{k}'},
\end{eqnarray}
and then one can pull out the derivatives resulting in the expression in Eq.~() of the paper.

\subsection{Calculation of the linear response transport coefficients}
The Hamiltonian is quadratic and $\vec{k}$ is a good quantum number (for $\psi=0$), therefore correlation functions can be calculated explicitly and in a concise form. The energies are
\begin{equation}
	\epsilon_{\pm,\vec{k}}=\pm E_\vec{k} =
			\sqrt{(M+\lambda^2 \vec{k}^2)^2+v^2\vec{k}^2},
\end{equation}
and we denote by $U_{\vec{k}}$ the matrix that diagonalizes the Hamiltonian $U_{\vec{k}} \hat{h}(\vec{k},\vec{k}) U^{\dagger}_{\vec{k}} = E_{\vec{k}}\sigma_z$. The Kubo correlation function between two operators of the type
\begin{equation}
	\hat{A} = \int\! \frac{d^2k d^2k'}{(2\pi)^2}\Psi^{\dagger'}_{\vec{k}}
	A_{\vec{k},\vec{k}'}\Psi_{\vec{k}'}
\end{equation}
are then
\begin{eqnarray}
	L_{\hat{A},\hat{B}}
	&=& iT
	\sum_{\alpha,\beta=\pm}
	\int\! 
	\frac{d^{2}\vec{k}d^{2}\vec{k}'}{(2\pi)^4}
	\left[
		U_{\vec{k}}A_{\vec{k},\vec{k}'}U^{\dagger}_{\vec{k}'}
	\right]_{\alpha,\beta}
	\frac{f(\epsilon_{\alpha,\vec{k}})-f(\epsilon_{\beta,\vec{k}'})}
	{(\epsilon_{\alpha,\vec{k}}-\epsilon_{\beta,\vec{k}'})
	(\epsilon_{\alpha,\vec{k}}-\epsilon_{\beta,\vec{k}'}+i\eta)}
	\left[
		U_{\vec{k}'}B_{\vec{k}',\vec{k}}U^{\dagger}_{\vec{k}}
	\right]_{\beta,\alpha},
\end{eqnarray}
and plugging into it $\vec{j}^{Q,N}_{\vec{q}=0}$ we get the Kubo contribution to the thermal or charge conductivity.

The energy magnetization is defined via the differential equation
\begin{equation}
	2\vec{M}^Q - T \frac{\partial \vec{M}^Q}{\partial T} = 
	-\frac{i}{2T}\vec{\nabla}_\vec{q} \times
	\langle \hat{h}_{-\vec{q}} ; \hat{\vec{j}}^Q_\vec{q} \rangle
	|_{\vec{q}\to 0},
\end{equation}
which we can recast as
\begin{equation}
	\frac{\partial}{\partial T}\left(\frac{\vec{M}^Q}{T^2}
	\right) = 
	\frac{i}{2T^4}\vec{\nabla}_\vec{q} \times
	\langle \hat{h}_{-\vec{q}} ; \hat{\vec{j}}^Q_\vec{q} \rangle
	|_{\vec{q}\to 0}.
\end{equation}
And the correlation here is given by
\begin{equation}
	\langle \hat{h}_{-\vec{q}} ; \hat{\vec{j}}^Q_\vec{q} \rangle = 
	T\sum_{\alpha,\beta=\pm}
	\int\! \frac{d^2\vec{k}}{(2\pi)^2}
	\left[
	U_{\vec{k}}h_{\vec{k},\vec{k}-\vec{q}}U^{\dagger}_{\vec{k}-\vec{q}}
	\right]_{\alpha,\beta}
	\frac{f(\epsilon_{\alpha,\vec{k}})-f(\epsilon_{\beta,\vec{k}-\vec{q}})}
	{\epsilon_{\alpha,\vec{k}}-\epsilon_{\beta,\vec{k}-\vec{q}}}
	\left[
		U_{\vec{k}-\vec{q}}\vec{j}^Q_{\vec{k}-\vec{q},\vec{k}}
		U^{\dagger}_{\vec{k}}
	\right]_{\beta,\alpha}.
\end{equation}

\subsection{Derivation of the lattice heat current}
The process on deriving the heat current on the lattice is similar to the derivation of the current in the continuum model, with the following required adjustments. The first one is that the heat current is defined on the links between sites $j^Q_x(x+a/2,y)$ and $j^Q_y(x,y+a/2)$, and also is not strictly local but has a finite support from adjacent sites. Consequently, the scaling with the gravitational potential, which is defined on the sites themselves, is not completely local. Rather, the different components of $j^Q_y(x,y+a/2)$ should scale like $[1+\psi(x,y)][1+\psi(x\pm a,y\pm a)]$ depending on the participating sites.

The continuity equation on the lattice is
\begin{equation}
	j^Q_x\left(x+\tfrac{a}{2},y\right)-j^Q_x\left(x-\tfrac{a}{2},y\right)
	+
	j^Q_y\left(x,y+\tfrac{a}{2}\right)-j^Q_x\left(x,y-\tfrac{a}{2}\right)
	= i a \left[1+\psi(\vec{r})\right]
	\left[\hat{h}_{\rm lat}(\vec{r}),\mathcal{H}_{{\rm lat},\psi}\right].
\end{equation}
This equation cannot uniquely define the current as we can add a divergence-free term $\vec{j}^Q \to \vec{j}^Q+\vec{g}$ where
\begin{eqnarray}
	a g_x(x,y) &=& f\left(x,y+\tfrac{a}{2}\right)-f\left(x,y-\tfrac{a}{2}\right),
	\nonumber \\ 
	a g_y(x,y) &=& -f\left(x+\tfrac{a}{2},y\right)+f\left(x-\tfrac{a}{2},y\right).
\end{eqnarray}
However combining this equation with the scaling requirement removes this ambiguity and allows us to derive the heat current on the lattice. The current in the $y$-direction is given by
\begin{equation}
	\hat{j}^Q_y(\vec{r}) = -i\frac{a}{N_x N_y}\sum_{\vec{k},\vec{k}'}
	e^{-i(\vec{k}-\vec{k}')\vec{r}}\Psi^{\dagger}_{\vec{k}}
	j^Q_{y,\vec{k},\vec{k}'}\Psi_{\vec{k}'}
\end{equation}
where
\begin{eqnarray}
	j^Q_{y,\vec{k},\vec{k}'} &=& \frac{i}{a^2}
	\left[v^2 e^{-i\frac{a}{2}(k'_y-k_y)}
	\cos\left(\frac{k'_y+k_y}{2}a\right)+M\lambda^2
	\right]	
	\sin\left(\frac{k_y+k'_y}{2}a\right)
	\nonumber \\ &&
	+\frac{\lambda^4}{4a^4}
	\left[2i e^{-i\frac{a}{2}(k'_y-k_y)} \sin(k_y a + k'_y a)
	+e^{i k_x a+i\frac{a}{2}(k_y+k'_y)}
	-e^{-i k'_x a-i\frac{a}{2}(k_y+k'_y)}
	\right]
	\nonumber \\ &&
	+\left[i\frac{v\lambda}{2a^2}
	e^{-i\frac{a}{2}(k'_x-k_x}
	\sin\left( \frac{k_x+k'_x+k_y+k'_y}{2}a \right)-i\frac{v M}{a}
	\sin\left( \frac{k_y+k'_y}{2}a	\right)
	\right]\sigma_x
	\nonumber \\ &&
	+i\frac{v\lambda^2}{2a^3}e^{-i\frac{a}{2}(k'_x-k_x)}
	\sin\left(\frac{k_x+k'_x+k_y+k'_y}{2}a\right)\sigma_y
	-i\frac{v^2}{2a^2}e^{-i\frac{a}{2}(k'_x-k_x)}
	\cos\left(\frac{k_x+k'_x+k_y+k'_y}{2}a\right)\sigma_z
	\nonumber \\ &&
	-\frac{1}{4}
	\bigg[e^{i\frac{a}{2}\left(k_x+k'_x-k_y-k'_y\right)}
	\left( \frac{\lambda^4}{a^4}
	+\frac{v\lambda^2}{a^3}(\sigma_x-\sigma_y)-i\frac{v^2}{a^2}\sigma_z\right)
	\nonumber \\ && -
	e^{-i\frac{a}{2}\left(k_x+k'_x-k_y-k'_y\right)}
	\left( \frac{\lambda^4}{a^4}+\frac{v\lambda^2}{a^3}(\sigma_x-\sigma_y)
	+i\frac{v^2}{a^2}\sigma_z\right)
	\bigg]
	\cos\left( \frac{k_x-k'_x}{2}a \right).
\end{eqnarray}

\end{widetext}

\end{document}